\documentclass[12pt]{article}

\usepackage{amssymb}
\usepackage{amsmath}
\usepackage{graphicx}
\usepackage{float}
\usepackage[margin=1in]{geometry}
\usepackage{verbatim}
\usepackage[usenames, dvipsnames]{color}

\begin{document}

\large
\centerline{\bf Elastic Scattering of Twisted Photons with Atomic Hydrogen}

\normalsize

\vskip 2em

\centerline{Jack Gallahan\footnote{jrgallahan11@gmail.com}}

\centerline{1129 Scherer Way, Osprey, FL}

\centerline{September 12, 2019}
\vskip 3em

\textbf{Abstract:} 
The previously derived vortex atomic form factor, which is directly related to a differential reaction cross section, is used to analyze the elastic scattering of twisted vortex photons with a hydrogenic atomic target. The vortex atomic form factor is expressed in a unified spherical basis and implemented in a MatLab code that numerically evaluates it using globally adaptive quadrature. The results of this code show the influence of variation in the photon wavelength, Rayleigh range, and scattering angle on differential reaction cross sections and the twist factor, which measures the impact of introducing orbital angular momentum. The recently suggested double mirror effect that accounts for a non-zero effect in the forward direction for twisted photon interactions is numerically confirmed. Finally, it is shown that differential reaction cross sections are greatly amplified when the Rayleigh range and photon wavelength are brought close to the scale of an atom. Experimental considerations and applications are briefly discussed, including quantum information, in which the scattering of twisted photons on atomic targets can be used to transfer information between light and matter. 

\newpage

\section{Introduction}

Twisted photons, which differ from their plane-wave counterparts in that they carry orbital angular momentum, offer more degrees of freedom than plane wave photons, giving them greater power to store information and macroscopically influence microscopic behavior [1-3]. Twisted electron, proton, and even atom beams have also been generated [4-6]. Here, twisted photons formulated in a Lageurre-Gaussian basis are considered. These are characterized by the parameters $(z_R, \lambda, p, l, m_s)$, whereas plane wave photons are described by only $(\lambda, m_s)$. $z_R$ represents the continuously variable and macroscopically adjustable Rayleigh range of the twisted photon, which measures the distance along the beam axis over which the cross sectional area of the beam in doubled. $l$ is the orbital angular momentum quantum number. Twisted photons carry discrete amounts of orbital angular momentum given by $l\hbar$ for integer values of $l$. Lageurre-Gaussian photons that carry orbital angular momentum $l$ include a helical phase factor $e^{i l    \phi}$ corresponding to rotation in the plane perpendicular to the direction of propagation. $p$ represents the number of nodes in the photonic radial distribution.

A wide variety of applications for twisted light have been explored including imaging [7-11], the construction of optical tweezers [12-15], and quantum information [16-25], where interactions between twisted photons and atomic targets have been investigated as a means of transferring information [26-29]. In these situations, it may be desirable to maximize the probability that a certain reaction occurs by varying experimentally adjustable parameters. With this in mind, the effect of three of these parameters, the scattering angle, Rayleigh range, and photon wavelength, is examined with an emphasis on determining the conditions that produce large differential reaction cross sections for the elastic scattering of twisted photons. In light of recent research on the production of vortex beams in the extreme ultraviolet and x-ray regions [30-36], interactions in the regime in which the photon wavelength is comparable to the scale of an atom are considered. The effect of shrinking the Rayleigh range to the atomic scale is also considered, although this is likely ahead of experiments.

This research follows the work of McGuire et. al. in two previous, closely related papers [2, 3]. The first derived a novel expression for the matrix element associated with the scattering of a Lageurre-Gauss beam with a hydrogen atom, called the vortex atomic form factor $M_v$, which is analogous to the commonly used plane wave form factor $M$. The second determined the normalization coefficient for this form factor that allowed it to converge to the plane wave limit as the Rayleigh range approaches infinity, and addressed the effects of a non-zero distance of closest approach, $\vec{b}$, between the center of the beam and the target atom. This research uses the vortex atomic form factor, which can be multiplied by the differential cross section for classical Thomson scattering to yield a quantum mechanical cross section, as the basis for its analysis. Afanasev et. al. [1] have used a Bessel function basis to avoid the paraxial approximation [37], which assumes that the beam envelope varies slowly along the beam axis and is only valid for cone angles less than about $\frac{\pi}{6}$ radians from the beam axis in each direction. However, it is taken here for mathematical simplicity.

In the following mathematical methods section, the four distinct coordinate systems used by McGuire et. al. [2, 3] are unified in a single mathematical basis to make numerical evaluation possible. In the results section, the effects of variation in the scattering angle $\Theta$, Rayleigh range $z_R$,  and photon wavelength $\lambda$ are shown under selected experimental conditions. The twist factor $T_v = \frac{|M_v|^2 - |M|^2}{|M|^2}$ [3] captures the effect orbital angular momentum has on the form factors. In the discussion section, experimental considerations and applications in quantum information are briefly considered. 

\section{Mathematical Methods} 

The atomic form factor for the interaction of plane wave beams with an atomic target, which is widely used to describe photoabsorption, photoemission, and photon scattering, is given by [2]
\begin{equation}
M = \int d\vec{r} \phi^*_{N_f, L_f, M_f}(\vec{r}) e^{-i\vec{k_f} \cdot \vec{r}}e^{i\vec{k_i} \cdot \vec{r}}\phi_{N_i, L_i, M_i}(\vec{r}),
\end{equation}
where $\phi_{N_i, L_i, M_i}$ and $\phi_{N_f, L_f, M_f}$ are the initial and final electronic states, and $\vec{k_i}$ and $\vec{k_f}$ are the initial and final angular wave numbers of the photon. The vortex atomic form factor is given by [3]
\begin{equation}
M_v = \frac{1}{2} \lambda z_R \int d\vec{r} \phi^*_{N_f, L_f, M_f}(\vec{r}) u^*_{p_f, l_f}(\vec{r})e^{-i\vec{k_f} \cdot \vec{r}} e^{i\vec{k_i} \cdot \vec{r}} u_{p_i, l_i}(\vec{r}) \phi_{N_i, L_i, M_i}(\vec{r}),
\end{equation}
where $\lambda$ is the wavelength of both the incoming and outgoing photons, $z_R$ is the Rayleigh range, $\phi_{N_i, L_i, M_i}$ and $\phi_{N_f, L_f, M_f}$ are the initial and final electronic states, and the plane wave factors $e^{i \vec{k_i} \vec{r}}$ and $e^{i \vec{k_f} \vec{r}}$ are modified by Laguerre-Gauss profile factors $u_{p_i, l_i}$ and $u_{p_f, l_f}$ that give the vector potential for twisted photons within the paraxial approximation [37]. In cylindrical coordinates, the Laguerre-Gauss factors are given by [37,38]
\begin{equation}
\begin{split}
& u_{p, l}(\rho, z, \phi) = \sqrt{\frac{2p!}{\pi(p + |l|)!}} \frac{1}{w(z)} \left[\frac{\rho\sqrt(2)}{w(z)}\right]^{|l|} exp\left[\frac{-\rho^2}{w(z)^2}\right] L^{|l|}_p\left(\frac{2\rho^2}{w(z)^2}\right) \\
& \times \ exp[il\phi] \ exp\left[\frac{ik\rho^2z}{2(z^2 + z^2_R)}\right] exp\left[-i(2p + |l| + 1)arctan\left(\frac{z}{z_R}\right)\right],
\end{split}
\end{equation}
where the twisted photon's momentum is in the $z$ direction, $w(z) = \sqrt{\frac{\lambda z_R}{\pi}(1+ \frac{z^2}{z^2_R})}$ gives the radius at which the beam intensity has decreased by a factor of $\frac{1}{e^2}$ from its maximum at the beam center, $p$ is the radial quantum number, $l$ is the azimuthal index or orbital angular momentum quantum number, and $L^{|l|}_p(x)$ is an associated Laguerre polynomial. $M_v$ is related to a differential reaction cross section for Compton scattering by [39, 40]
\begin{equation}
\frac{d\sigma_C}{d\Omega} = |M_v|^2 \frac{\omega_f}{\omega_i} r_0^2 |\hat{\Lambda_i} \cdot \hat{\Lambda_f}|^2,
\end{equation}
where $\hat{\Lambda_i}$ and $\hat{\Lambda_f}$ are the polarization directions of the incoming and outgoing photons, $r_0 = 2.818 \times 10^{-15}$ m is the classical radius of an electron, and $\omega_i$ and $\omega_f$ are the angular frequencies of the incoming and outgoing photons, which, for the purposes of this paper, are equal. The Klein-Nishina relativistic correction term is omitted. 

The principal mathematical obstacle in numerically evaluating this form factor is the presence of four distinct coordinate systems: two spherical ones $(r, \theta, \phi)$ and $(r, \theta', \phi')$ for the initial and final electronic states and two cylindrical ones $(\rho, z, \phi)$ and $(\rho', z', \phi')$ for the initial and final Laguerre-Gauss factors. The twisted photon propagates in the $\theta = 0$, $\theta' = 0$, $\rho = 0$, and $\rho' = 0$ directions for each respective coordinate system. The final spherical basis is rotated from the initial by the scattering angle $\Theta$.

Due to the presence of hydrogenic wave functions, it is only convenient to calculate the form factor in spherical coordinates, eliminating the possibility of using a common Cartesian grid, as was suggested by McGuire et. al. [3]. In light of this, all coordinates are expressed in terms of the initial spherical basis. The initial and final cylindrical coordinate systems can easily be converted to their respective spherical coordinate system with the simple relations $\rho = rsin(\theta)$ and $z = rcos(\theta)$. However, expressing the final spherical basis in terms of the initial spherical basis poses a greater mathematical challenge. In order to map the rotated spherical basis onto the original one, two intermediate Cartesian grids will be used as shown below:
\begin{figure}[H]
\begin{center}
\includegraphics[scale= 1]{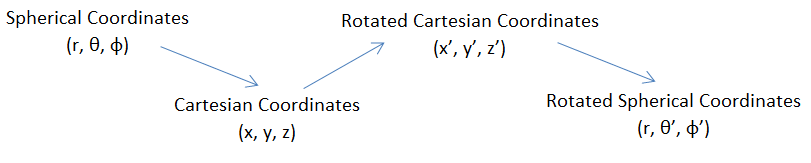}
\end{center}   
  \caption{Mathematical process for expressing rotated spherical basis in terms of the original spherical basis.}
\end{figure}
Here the $yz$ plane in the first Cartesian grid corresponds to the $\phi = \frac{\pi}{2}$ plane in the first spherical coordinate system. The rotation will be executed about the x-axis of the Cartesian grid in the direction such that the positive y-axis is rotated toward the positive z-axis, although either direction would yield the same final answer. The first conversion in Figure 1 is accomplished by the well-known formulas
\begin{equation}
\begin{split}
x & = rsin(\theta)cos(\phi), \\
y & = rsin(\theta)sin(\phi), \\
z & = rcos(\theta).
\end{split}
\end{equation}
Since the rotation of the Cartesian grid is about the x-axis, $x' = x$. Considering the points $(y, z)$ and $(y', z')$ in the $yz$ plane, 
\begin{figure}[H]
\begin{center}
\includegraphics[scale= .6]{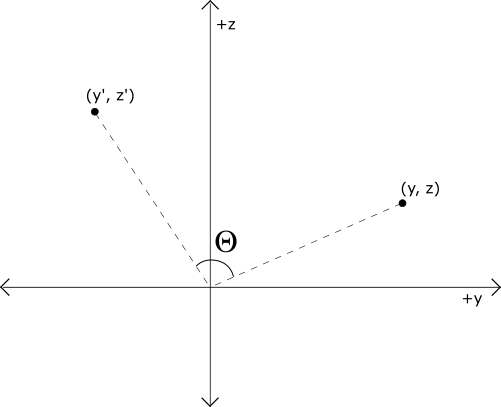}
\end{center}   
  \caption{Rotation by $\Theta$ in the yz plane of scattering.}
\end{figure}
\noindent the angle the position vector $\langle y', z' \rangle$ makes with the positive y-axis is clearly $\Theta \ + \  arctan\left(\frac{z}{y}\right)$. The distance from either point to the origin is $\sqrt{y^{2} + x^{2}}$. It follows that 
\begin{equation}
\begin{split}
y' & = \sqrt{y^{2} + x^{2}}cos\left(\Theta \ + \  arctan\left(\frac{z}{y}\right)\right),\\
z' & = \sqrt{y^{2} + x^{2}}sin\left(\Theta \ + \  arctan\left(\frac{z}{y}\right)\right).
\end{split}
\end{equation}
Combining Equation 5 and Equation 6, the rotated Cartesian coordinates can be expressed in terms of the original spherical basis as follows:
\begin{equation}
\begin{split}
x' & = rsin(\theta)cos(\phi),\\
y' & = r\sqrt{sin^2(\theta)sin^2(\phi) + cos^2(\theta)}cos\left(\Theta + acrtan\left(cot(\theta)csc(\phi)\right)\right),\\
z' & = r\sqrt{sin^2(\theta)sin^2(\phi) + cos^2(\theta)}sin\left(\Theta + acrtan\left(cot(\theta)csc(\phi)\right)\right).
\end{split}
\end{equation}

Since only rotation is involved, $r' = r$. The rotated spherical basis is related to the rotated Cartesian grid by the common conversion formulas
\begin{equation}
\begin{split}
\theta' & = arccos\left(\frac{z'}{r}\right),\\
\phi' & = arctan\left(\frac{y'}{x'}\right).
\end{split}
\end{equation}
Substituting Equation 7 yields 
\begin{equation}
\begin{split}
\theta' & = arccos\left(\sqrt{sin^2(\theta)sin^2(\phi) + cos^2(\theta)}sin\left(\Theta + acrtan\left(cot(\theta)csc(\phi)\right)\right)\right),\\
\phi'& = arctan\left(\frac{\sqrt{sin^2(\theta)sin^2(\phi) + cos^2(\theta)}cos\left(\Theta + acrtan\left(cot(\theta)csc(\phi)\right)\right)}{sin(\theta)cos(\phi)}\right),
\end{split}
\end{equation}
completing the transformation.

\section{Results}

Equations 2, 3, 4 and 9 were integrated into a MatLab code that numerically evaluates the differential reaction cross section for elastic scattering with plane wave photons, the differential reaction cross section for elastic scattering with twisted photons, and the twist factor $T_v$ using MatLab's "integral3" function, a globally adaptive quadrature routine. First, the effect of varying the scattering angle $\Theta$ on the twist factor $T_v$ is examined. 

\begin{figure}[H]
\begin{center}
\includegraphics[scale= .4]{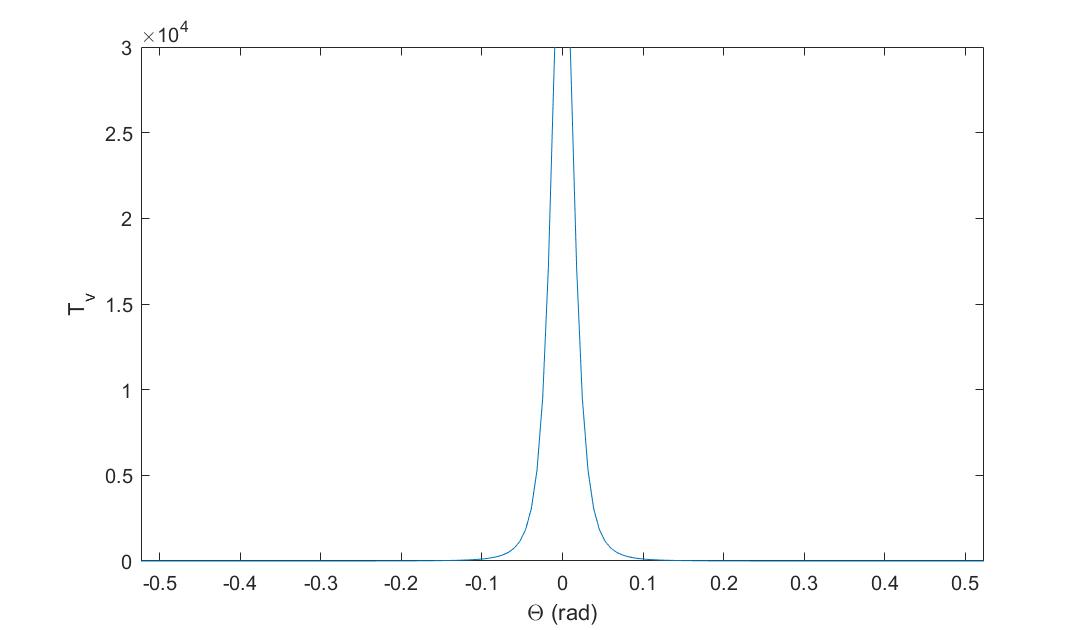}
\end{center}   
  \caption{Twist factor $T_v$ versus scattering angle $\Theta$ for a photon of wavelength $550$ nm, a Rayleigh range of $1000$ nm, an electronic transition from 1s to 3d, and a transfer of 2 units of orbital angular momentum such that $l_i = 1$, $l_f = -1$, and $p_i = p_f = 0$. The divergence at $\Theta = 0$ confirms a double mirror effect, as explained in the text.}
\end{figure}

In Figure 3, the twist factor is shown, mathematically defined as $T_v = \frac{|M_v|^2 - |M|^2}{|M|^2}$ and representing the effect of introducing orbital angular momentum on scattering. The singularity at $\Theta = 0$ in Figure 3 shows that for plane wave interactions, elastic scattering is forbidden in the forward direction. For twisted photons, on the other hand, the differential cross section for elastic scattering is non-zero in the forward direction. This result confirms the existence of the double mirror effect suggested by McGuire et. al. in [2] that accounts for a non-zero effect for twisted photon interactions in the forward direction. Parity blocks plane wave photon scattering in the forward direction, but the $e^{il\phi}$ term present in the spacial distribution of Laguerre-Gauss beams removes this restriction in the same way that right handedness is restored in the reflection of a second mirror. Next, the influence of the Rayleigh range on scattering is analyzed for two electronic transitions.

\begin{figure}[H]
\begin{center}
\includegraphics[scale = 0.30]{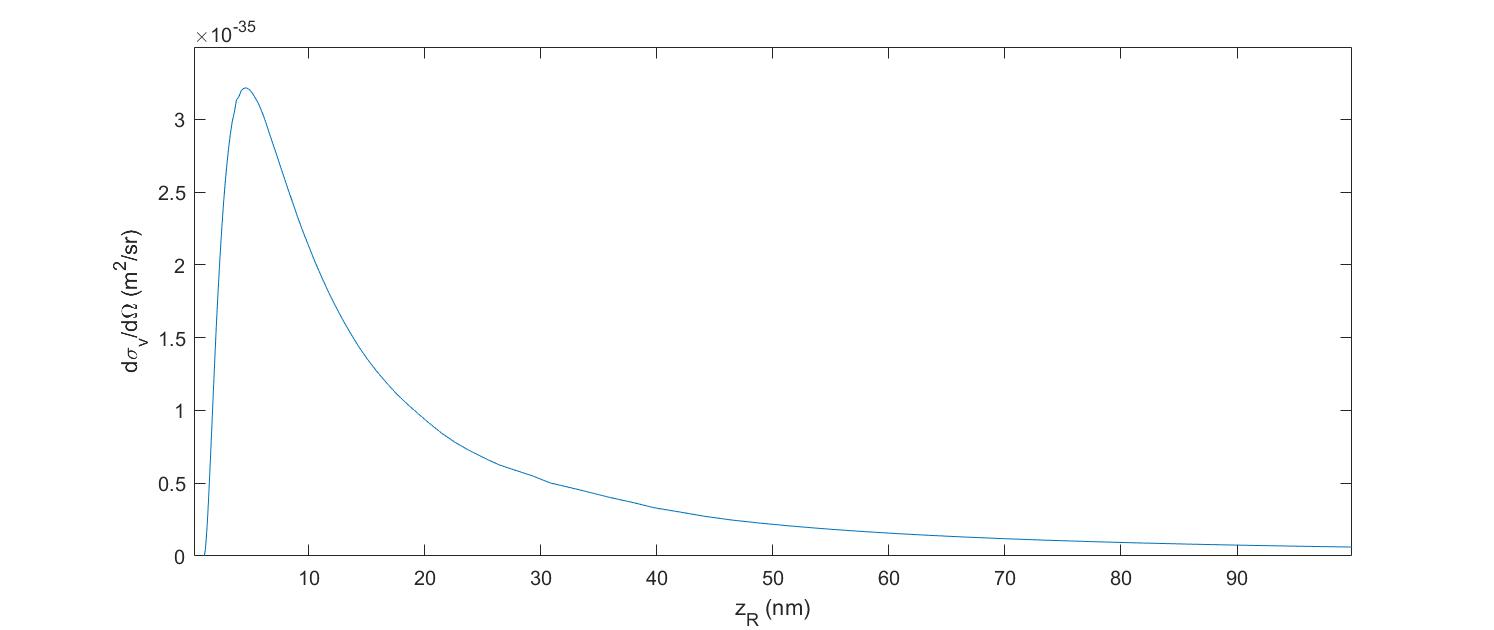}
\includegraphics[scale = 0.30]{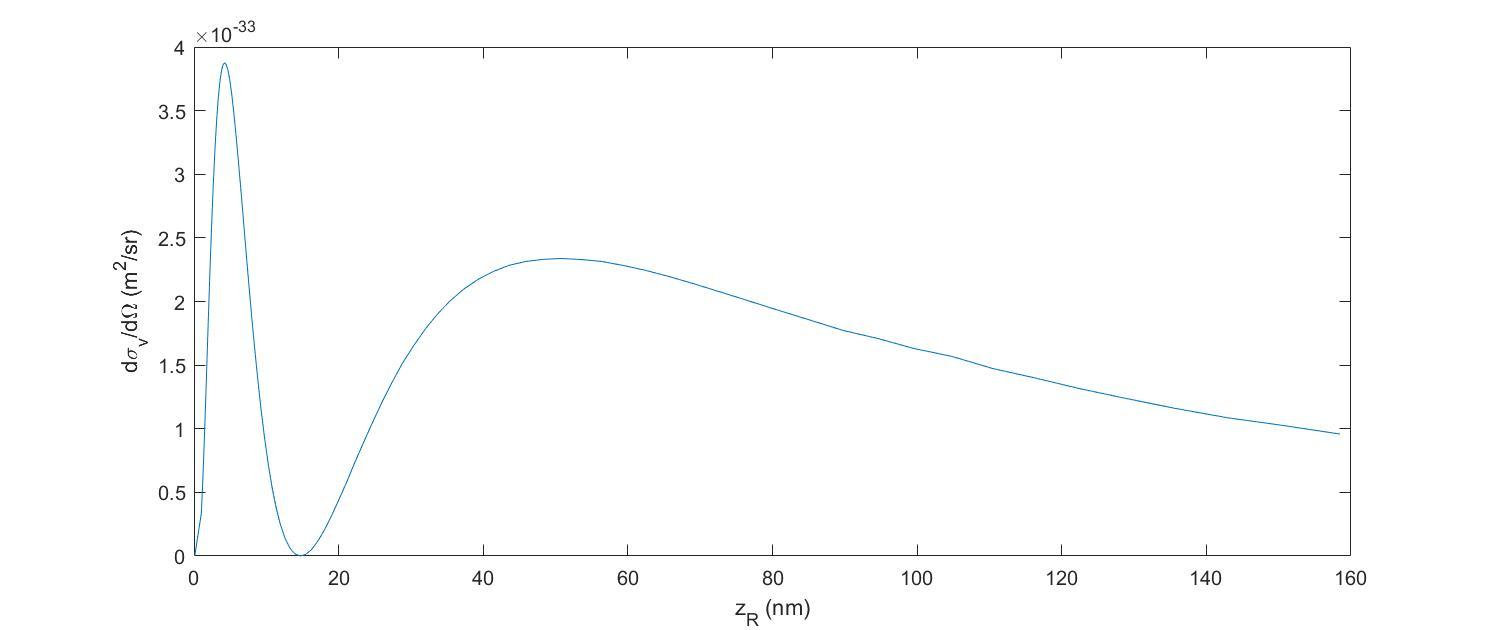}
\end{center}   
  \caption{Top: differential reaction cross section for the elastic scattering of a twisted photon with atomic hydrogen versus Rayleigh range $z_R$ for a photon with wavelength $0.1$ nm, a scattering angle of $0$, an electronic transition from 1s to 3d, and a transfer of 2 units of orbital angular momentum such that $l_i = 1$, $l_f = -1$, and $p_i = p_f = 0$. Bottom: same axes and parameters as top graph, but with an electronic transition of 3s to 3d.}
\end{figure}

For any electronic transition, as the Rayleigh range approaches infinity, the twisted photon becomes indistinguishable from a plane wave photon [2, 3]. As a result, the vortex atomic form factor should approach the value of the atomic form factor as $z_R$ increases without bound. Since scattering for plane wave photons is forbidden by parity in the forward direction, this means that the vortex atomic form factor and, recalling Equation 4, the differential reaction cross section for twisted photons should approach $0$ as $z_R$ approaches infinity when $\Theta = 0$. Both the 1s to 3d and 3s to 3d transitions meet this condition.

For the 1s to 3d transition, the vortex atomic form factor $M_v$ and the differential reaction cross section $\frac{d\sigma_v}{d\Omega}$ are amplified when $z_R$ is small. This is also generally true of the 3s to 3d transition, although the effect vanishes when the wavelength is about thirty times the atomic radius. Under typical experimental conditions, the atomic radius, photon wavelength, and Rayleigh range occupy vastly different length scales, with $a_0 << \lambda << z_R$. The results in Figure 4 and Figure 5 show that deviating from this regime yields more detailed and pronounced effects. In fact, when $z_R$ is on the order of $1$ mm, the differential cross section is so small that, in an experiment, it would likely be washed out by the slightly non-forward scattering of marginally diverging light. Figure 4 shows the existence of one maximum for the 1s to 3d transition and two maxima for the 3s to 3d transition. At these points, the differential cross section and therefore experimental reaction count rate for elastic scattering is the largest. Finally, the impact of varying the photon wavelength on the differential reaction cross section for a twisted photon interaction is considered. 

\begin{figure}[H]
\begin{center}
\includegraphics[scale= .30]{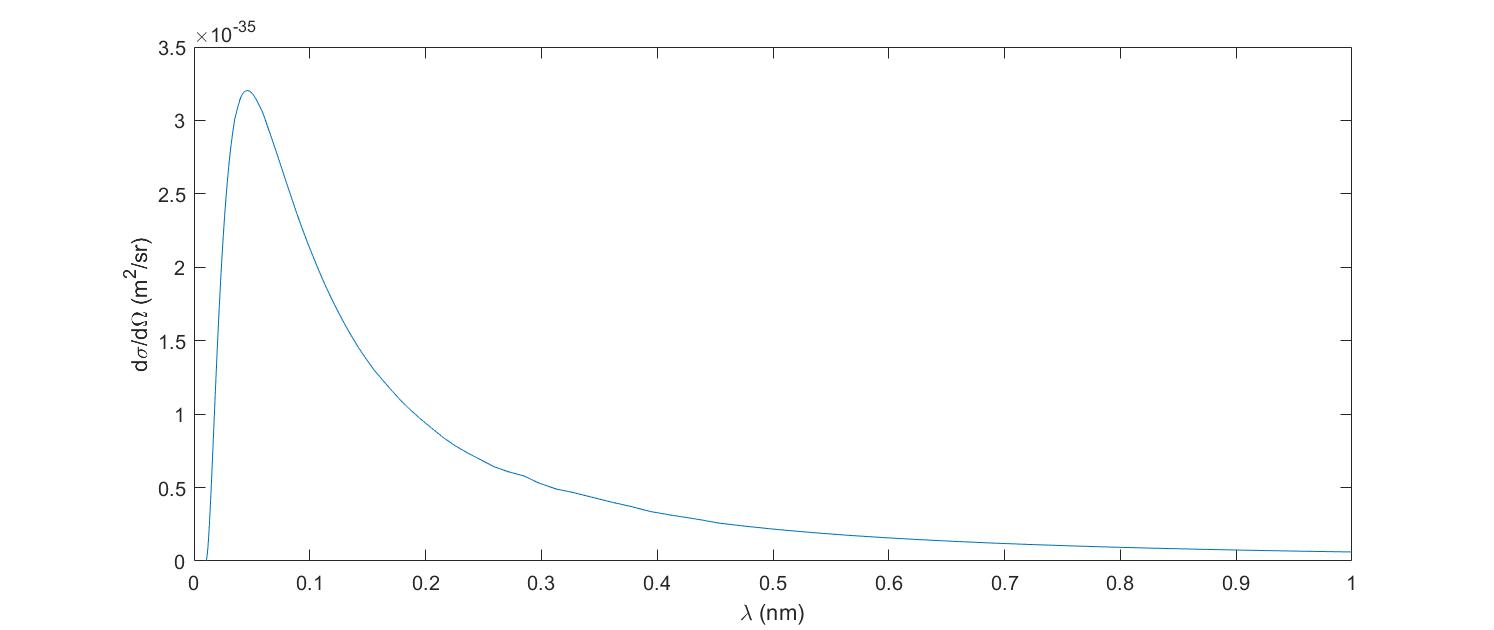}
\includegraphics[scale= .30]{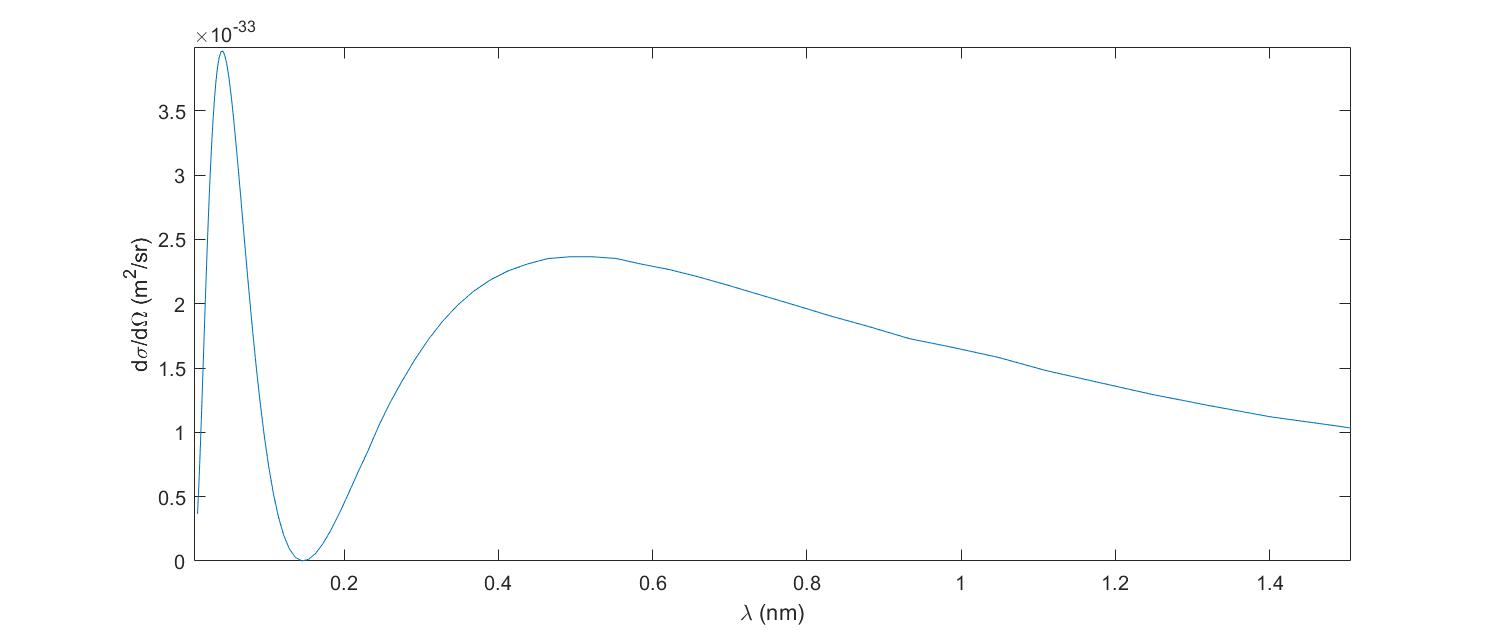}
\end{center}   
  \caption{Top: differential reaction cross section for the elastic scattering of a twisted photon with atomic hydrogen versus wavelength $\lambda$ for a photon with a Rayleigh range of $10$ nm, a scattering angle of $0$, and a transfer of 2 units of orbital angular momentum such that $l_i = 1$, $l_f = -1$, and $p_i = p_f = 0$. Bottom: same axes and parameters as the top graph, but with a 3s to 3d electronic transition.}
\end{figure}

Figure 5 shows that when the twisted photon wavelength $\lambda$ is close to the size of a hydrogen atom (about $0.12$ nm), the differential cross section for elastic scattering is generally significantly larger than it is when the wavelength $\lambda$ is in the visible spectrum, for example. As mentioned above for Rayleigh range, wavelengths much greater than the size of a hydrogen atom yield effects that are so minuscule that they would be unlikely to survive in a slightly divergent or convergent beam. Previous research has made it increasingly possible to generate twisted photons in the x-ray and extreme ultraviolet regions [22-28], opening the door to experiments in which $\lambda$ and $a_0$ are comparable rather than disparate. Finally, it is of interest that the variation in the two adjustable length scales involved in the scattering process---the Rayleigh range $z_R$ and the photon wavelength $\lambda$---produce nearly identical behavior in the differential cross section. 

\section{Discussion}

The vortex factor $T_v$ is particularly relevant to experiments for two reasons. First, $T_v$ may be used to transform data for plane-wave interactions into those for twisted vortex beam interactions. In particular, the quantity $T_v + 1 =  \frac{|M_v|^2}{|M|^2}$ can serve as a vortex conversion factor [3]. Second, the fact that $T_v$ is a ratio of form factors could lead errors due to fluctuations in experimental setup to cancel, making it a robust link from theory to experiment. The Rayleigh range also has interesting implications. In principal, one could influence behavior on the atomic scale by macroscopically adjusting the cone angle of the beam like a pair of tweezers, coupling the atomic and macroscopic scales. Finally, geometric structure factors [41], may be used to convert the data for single atoms presented in this research to data for molecules or crystals. 

One of the major applications of the interaction of twisted vortex photons with atomic targets is quantum information. The presence of the five experimentally adjustable control parameters $(z_R, \lambda, p, l, m_s )$, three of which are continuously variable, enables more information to be encoded in twisted photons that in plane wave photons.  In particular, the orbital angular momentum quantum number holds great potential to store information, since its value is discrete and unbounded. Since the specific configuration of the light beam influences its effect on an atomic target, the information stored in twisted light can be transferred to matter through scattering [26-29]. More broadly, the additional degrees of freedom present in twisted vortex photons presents the possibility of creating high-dimensional qudits (or quantum bits with more that two available states) that may be used in quantum cloning, quantum communication, explorations of large violations of local-realistic theories, and quantum computation [24]. 

\section{Summary}

The multiple coordinate systems present in the original vortex atomic form factor were unified into a single spherical basis. A MatLab code was created to numerically evaluate the atomic form factor, vortex atomic form factor, their related differential reaction cross sections, and the twist factor, which captures the effect of orbital angular momentum on scattering. Theoretical results were then presented focusing on the influence of the scattering angle, the Rayleigh range, and the photon wavelength on the elastic scattering of twisted photons with atomic hydrogen. The results confirmed the double mirror effect noted by McGuire et. al. [30] which explains the presence of a non-zero differential reaction cross section for twisted photon interactions in the forward direction while plane wave scattering is forbidden by parity in the forward direction. In addition, it was demonstrated that cross sections and reaction count rates are dramatically amplified and more complex in the regime in which the Rayleigh range and photon wavelength are comparable to the size of an atom. Finally, experimental considerations and possible applications in quantum information were touched upon. 

\section{Acknowledgments}

The author gratefully acknowledges useful exchanges with J. H. McGuire and A. Afanasev.




\end{document}